\documentclass[aps,prc,showpacs,twocolumn,nofootinbib,groupedaddress,preprintnumbers]{revtex4-1}

\usepackage{calligra}
\usepackage{longtable}
\usepackage{graphicx}
\usepackage{dcolumn}
\usepackage{bm}
\usepackage{amsmath,amssymb}
\usepackage{longtable}

\newcommand{\bra}[1]{\langle \, #1 \, |}
\newcommand{\ket}[1]{| \, #1 \, \rangle}
\newcommand{\kket}[1]{\, #1 \, \rangle}

\newcommand{\re}{\text{Re }}
\newcommand{\im}{\text{Im }}
\newcommand{\Jost}{\text{\textcalligra{f}\;}}

\begin{document}

\preprint{YITP-14-55}


\title{Hadron mass scaling near the $s$-wave threshold}


\author{Tetsuo~Hyodo}
\email[]{hyodo@yukawa.kyoto-u.ac.jp}
\affiliation{Yukawa Institute for Theoretical Physics, Kyoto University, Kyoto 606-8502, Japan}


\date{\today}

\begin{abstract}
The influence of a two-hadron threshold is studied for the hadron mass scaling with respect to some quantum chromodynamics parameters. A quantum mechanical model is introduced to describe the system with a one-body bare state coupled with a single elastic two-body scattering. The general behavior of the energy of the bound and resonance state near the two-body threshold for a local potential is derived from the expansion of the Jost function around the threshold. It is shown that the same scaling holds for the nonlocal potential induced by the coupling to a bare state. In $p$ or higher partial waves, the scaling law of the stable bound state continues across the threshold describing the real part of the resonance energy. In contrast, the leading contribution of the scaling is forbidden by the nonperturbative dynamics near the $s$-wave threshold. As a consequence, the bound state energy is not continuously connected to the real part of the resonance energy. This universal behavior originates in the vanishing of the field renormalization constant of the zero-energy resonance in the $s$ wave. A proof is given for the vanishing of the field renormalization constant, together with a detailed discussion.
\end{abstract}




\maketitle

\section{Introduction}

The properties of hadrons are determined by highly nonperturbative dynamics of quantum chromodynamics (QCD). In some cases, however, the mass of hadrons can be expressed by a systematic expansion of certain QCD parameters. The mass of hadrons with light quarks is expanded by the light quark mass $m_{q}$ in chiral perturbation theory~\cite{Scherer:2012xha}. When the hadron contains one heavy quark, its mass can be given in powers of the inverse of the heavy quark mass $1/m_{Q}$, with the leading contribution of $\mathcal{O}(m_{Q})$~\cite{Manohar:2000dt}. It is also possible to express the mass of hadrons in the $1/N_{c}$ expansion~\cite{Jenkins:1998wy}. These expansions dictate the scaling laws of the hadron mass as functions of the dimensionless parameter $x=m_{q}/\Lambda, \Lambda/m_{Q}$, and $1/N_{c}$ with $\Lambda$ being the nonperturbative energy scale of QCD. The leading contribution of the expansions is determined by the construction of hadrons. For instance, the mass of the Nambu-Goldstone bosons scales with $(m_{q}/\Lambda)^{1/2}$, while the expansion of other light hadrons contains a constant term as the leading contribution. In the heavy sector, the mass of the singly heavy hadrons primarily scales as $(\Lambda/m_{Q})^{-1}$ for the heavy quark mass and $(m_{q}/\Lambda)^{0}$ for the light quark mass. The ordinary mesons and ordinary baryons behave as $(1/N_{c})^{0}$ and $(1/N_{c})^{-1}$, respectively. The higher order corrections can also be calculated systematically. The mass scaling is useful, for instance, to extrapolate the results of the lattice QCD simulation to the physical quark mass region.

Because each hadron has its own scaling law, one may encounter the situation where a hadron mass goes across a two-hadron threshold with the same quantum numbers. For instance, the signal of the $H$ dibaryon is found as a stable bound state in the heavy $m_{q}$ region in lattice QCD~\cite{Inoue:2010es,Beane:2010hg}, while it is considered to become a resonance slightly above the $\Lambda\Lambda$ threshold at the physical point~\cite{Shanahan:2011su}. In fact, many hadron resonances are obtained as stable bound states in the heavy $m_{q}$ simulation, because the lowest two-hadron threshold usually contains the pion whose mass grows as $\sqrt{m_{q}}$. 

When the hadron mass moves across the threshold, one may naively expect that the same scaling law with the bound state energy describes the real part of the resonance energy above the threshold. It should be noted, however, that the threshold dynamics at the hadronic level is also highly nonperturbative~\cite{Braaten:2004rn}. For instance, the two-body scattering length in the $s$-wave channel diverges when the binding energy is sent to zero. Thus, the coupling effect to the two-hadron channel should be carefully examined in the threshold energy region.

\section{Formulation}

The mass scaling considered in the following is the case when a hadron (hereafter called the bare state) approaches a two-body threshold from the lower energy side. Only the lowest energy two-body threshold is taken into account among the possible states having the same quantum numbers with the bare state. This two-body channel will be referred to as the scattering channel. Thus, the analysis focuses on the elastic two-body scattering. The near-threshold kinematics is discussed where the nonrelativistic treatment is applicable. The absence of the long-range interaction is assumed. 

The effect of the scattering channel is described by the coupled-channel Hamiltonian~\cite{Baru:2003qq},
\begin{align}
    \begin{pmatrix}
    \hat{H}_{0} & \hat{V} \\
    \hat{V} & \hat{H}_{\rm sc}
    \end{pmatrix} 
    \ket{\Psi}
    &= 
    E\ket{\Psi},\quad
    \ket{\Psi}
    =
    \begin{pmatrix}
    c(E)\ket{\psi_{0}} \\
    \chi_{E}(\bm{p}) \ket{\bm{p}}
    \end{pmatrix} ,
    \label{eq:Hamiltonian}
\end{align}
where $\hat{H}_{0}$ ($\hat{H}_{\rm sc}$) is the Hamiltonian for the bare state (scattering) channel, $\hat{V}$ is the transition potential, and $c(E)$ [$\chi_{E}(\bm{p})$] is the wave function for the bare state (scattering) channel component $\ket{\psi_{0}}$ ($\ket{\bm{p}}$). In the scattering channel, the eigenvalue is $\hat{H}_{\rm sc} \ket{\bm{p}}=p^{2}/(2\mu) \ket{\bm{p}}$ with the reduced mass $\mu$.\footnote{Possible residual two-body potential $\hat{V}_{\rm sc}$ can be treated perturbatively by a proper field redefinition~\cite{Weinberg:1962hj,Weinberg:1963zz}.} For the bare state channel, the eigenvalue is given by $\hat{H}_{0}\ket{\psi_{0}}=M_{0}\ket{\psi_{0}}$ where $M_{0}$ is the energy of the bare state measured from the threshold in the absence of the scattering channel. It is considered that the scaling of $M_{0}(x)$ is known with respect to the QCD parameter $x$. The aim of this paper is to determine the scaling of the eigenenergy of the coupled-channel Hamiltonian $E_{h}(x)$. This enables one to relate the eigenenergy $E_{h}$ and the bare state energy $M_{0}$. 

First, consider the eigenenergy of the system~\eqref{eq:Hamiltonian} for a fixed $x$. To this end, the bound state channel is eliminated by the Feshbach method~\cite{Feshbach:1958nx,Feshbach:1962ut}. The effective potential which acts on the scattering channel $\ket{\bm{p}}$ is given by
\begin{align}
    \hat{V}_{\rm eff}(E)
    &= 
    \frac{\hat{V}\ket{\psi_{0}}
    \bra{\psi_{0}}\hat{V}}{E-M_{0}} .
    \label{eq:effectiveint}
\end{align}
By solving the Lippmann-Schwinger equation, the two-body scattering amplitude is obtained as
\begin{align}
    f(\bm{p},\bm{p}^{\prime},E)
    &= 
    -\frac{4\pi^{2}\mu \bra{\bm{p}}\hat{V}\ket{\psi_{0}}\bra{\psi_{0}}\hat{V}\ket{\bm{p}^{\prime}}}{E-M_{0}-\Sigma(E)} ,
    \label{eq:amplitude}
\end{align}
where the self-energy is defined as
\begin{align}
    \Sigma(E)
    &= \int \frac{\bra{\psi_{0}}\hat{V}\ket{\bm{q}}
    \bra{\bm{q}}\hat{V}\ket{\psi_{0}}}{E-q^{2}/(2\mu)+i0^{+}}d^{3}q .
    \label{eq:selfenergy}
\end{align}
The eigenenergy $E_{h}$ of the Hamiltonian is identified from the pole of the amplitude~\eqref{eq:amplitude}, namely,
\begin{align}
    E_{h}-M_{0}
    &= 
    \Sigma(E_{h})
    \label{eq:polecond} .
\end{align}
For a sufficiently large $|E_{h}|$, the self-energy behaves as $\Sigma(E_{h})\sim 1/E_{h}$ where the scattering state contribution is suppressed and the eigenenergy behaves as $E_{h}\sim M_{0}$. This means that the effect of the scattering channel is negligible in the energy region far away from the threshold, and the scaling of the eigenenergy $E_{h}(x)$ can be well described by the scaling of the bare mass $M_{0}(x)$, as naively expected. Nontrivial behavior emerges near the threshold.

\section{Threshold behavior from \\
the Jost function}\label{sec:nearthreshold}

To focus on the near-threshold phenomena, the QCD parameter $x$ is adjusted such that the eigenenergy appears exactly on top of the threshold $(E_{h}=0)$. This corresponds to setting the bare mass as $\bar{M}_{0}=-\Sigma(0)> 0$.\footnote{For a finite coupling of the scattering state and the bare state, $\Sigma(0)$ must be nonzero. The case $E_{h}=\bar{M}_{0}=0$, which corresponds to the vanishing of the coupling, will be separately discussed in Sec.~\ref{sec:decoupling}.} In this case, the scattering amplitude has a pole at zero energy. The pole of the amplitude is equivalent to the zero of the Jost function $\Jost_{l}(p)$ (Fredholm determinant) for the $l$th partial wave with the eigenmomentum $p=\sqrt{2\mu E_{h}}$. The properties of the Jost function are summarized in Appendix (see also Ref.~\cite{Taylor}). From the expansion of the Jost function around $p=0$ in Eq.~(A6), when $\Jost_{l}(p)=0$ at $p=0$, it can be expanded as
\begin{align}
    \Jost_{l}(p)
    &= 
    \begin{cases}
    i\gamma_{0}p + \mathcal{O}(p^{2}) & l=0 \\
    \beta_{l}p^{2} + \mathcal{O}(p^{3}) & l\neq 0 \\
    \end{cases} .
    \label{eq:Jost}
\end{align}
The real expansion coefficients $\gamma_{0}$ and $\beta_{l}$ are determined by the potential and the wave function. It is shown for a general local potential that $\Jost_{l}(p)$ goes to zero exactly as $p$ ($p^{2}$) for $l=0$ ($l\neq 0$)~\cite{JMP1.319} so that $\gamma_{0}$ and $\beta_{l}$ are guaranteed to be nonzero. This means that the zero of the Jost function at the threshold is simple for $l=0$, while it is double for $l\neq 0$. For the nonlocal potential~\eqref{eq:effectiveint}, the result of Ref.~\cite{JMP1.319} cannot be directly applied. It is nevertheless demonstrated in Secs.~\ref{sec:Zrelation} and \ref{sec:theorem} that the same scaling law is derived for the potential~\eqref{eq:effectiveint}, as long as the pole exists at the threshold.

The bare mass is then shifted as $\bar{M}_{0}\to \bar{M}_{0}+\delta M$ by changing the QCD parameter $x$ and examining the modification of the eigenenergy. For a given $\bar{M}_{0}$, it is always possible to consider a sufficiently small shift $\delta M\ll \bar{M}_{0}$. The effective potential at zero energy is then modified by
\begin{align}
    \hat{V}_{\rm eff}
    &\to
    \left(1+\frac{\delta M}{-\bar{M}_{0}}\right)
    \hat{V}_{\rm eff}
    \equiv \left(1+\delta \lambda\right)
    \hat{V}_{\rm eff} ,
    \label{eq:potentialmod}
\end{align}
where $\delta\lambda = -\delta M/\bar{M}_{0}$. Thus, the small shift of the bare mass results in the multiplicative modification of the strength of the effective potential. When the bare mass $\bar{M}_{0}$ is decreased (increased), the strength of the potential is enhanced by $\delta \lambda>0$ (reduced by $\delta\lambda<0$) and the formation of a bound (resonance) state is expected. For a positive $\delta\lambda$, the eigenmomentum in the leading order of $\delta\lambda$ is given by (see Appendix)
\begin{align}
    p&=
    i(\alpha_{0}^{\prime}/\gamma_{0}) \delta\lambda 
    \quad l=0 , \label{eq:momentum0}\\
    p^{2}
    &=
    -(\alpha_{l}^{\prime}/\beta_{l}) \delta\lambda 
    \quad l\neq 0 ,
    \label{eq:momentuml}
\end{align}
with $\alpha_{l}^{\prime}=d\alpha_{l}/d(\delta\lambda)|_{\delta\lambda=0}$. Thus, the energy of the bound state is 
\begin{align}
    E_{h}
    =&
    \begin{cases}
    -F_{0}\; \delta\lambda^{2}
    =-\tilde{F}_{0}\; \delta M^{2} & l=0 \\
    -F_{l}\; \delta\lambda 
    =\tilde{F}_{l}\; \delta M 
    & l\neq 0
    \end{cases} ,
    \quad 
    \delta\lambda > 0
    \label{eq:bindingenergy} ,
\end{align}
with the positive coefficients $F_{0}=(\alpha_{0}^{\prime})^{2}/(2\mu\gamma_{0}^{2})$, $F_{l}=\alpha_{l}^{\prime}/(2\mu\beta_{l})$ ($l\neq 0$), $\tilde{F}_{0}=F_{0}/\bar{M}_{0}^{2}$, and $\tilde{F}_{l}=F_{l}/\bar{M}_{0}$ ($l\neq 0$). It is found that, with a small increase of the potential strength by the factor $1+\delta\lambda$ (small decrease of the bare mass $\delta M$), the binding energy grows linearly in $\delta \lambda$ ($\delta M$) for $l\neq 0$ and quadratically for $l=0$. 

\begin{figure}[tbp]
    \centering
    \includegraphics[width=8cm,clip]{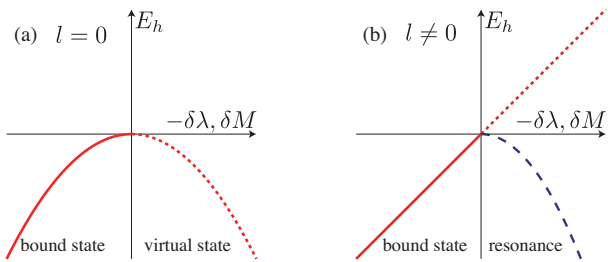}
    \caption{\label{fig:schematic}
    (Color online) Schematic illustration of the near-threshold 
    eigenenergy $E_{h}$ for $l=0$ (a) and $\l\neq 0$ (b) as a function of 
    the variation of the potential $\delta\lambda$ or the variation of the 
    bare mass $\delta M$. The solid lines represent the bound state 
    energy, the dotted lines stand for the real part of the resonance 
    energy ($l\neq 0$) and the energy of the virtual state ($l=0$), and 
    the dashed line represents the imaginary part of the resonance energy 
    ($l\neq 0$).}
\end{figure}

This result can be analytically continued to the negative $\delta\lambda$ region. For $l=0$, the eigenenergy is negative. This solution corresponds to the virtual state because the eigenmomentum has the opposite sign from the bound state. On the other hand, for $l\neq 0$, the eigenenergy becomes complex, so the pole represents the resonance solution. The real part is determined by the same formula with Eq.~\eqref{eq:bindingenergy}. The imaginary part comes from the higher order term $i\gamma_{l}p^{2l+1}$. To summarize, for $\delta\lambda<0$, the eigenenergy scales as
\begin{align}
    \begin{cases}
    E_{h}=-F_{0}\; \delta\lambda^{2}
    =-\tilde{F}_{0}\; \delta M^{2} & l=0 \\
    \re E_{h}
    =-F_{l}\; \delta\lambda 
    =\tilde{F}_{l}\; \delta M 
    & l\neq 0 
    \end{cases}
    , 
    \quad 
    \delta\lambda < 0
    \label{eq:eigenenergy} ,
\end{align}
and $\im E_{h}\propto (\delta\lambda)^{l+1/2}$ for $l\neq 0$. These behaviors are illustrated in Fig.~\ref{fig:schematic}. For $l\neq 0$, the scaling of the bound state energy continues above the threshold as the real part of the resonance energy. In the $s$-wave case, the bound state does not continuously turn into a resonance, but becomes a virtual state.

The near-threshold scaling can also be understood by the effective range expansion. The partial wave scattering amplitude is given by
\begin{align}
    f_{l}(p)
    =&
    \frac{p^{2l}}{-\frac{1}{a_{l}}+\frac{r_{l}}{2}p^{2}+\mathcal{O}(p^{4})-ip^{2l+1}} ,
\end{align}
where $a_{l}$ and $r_{l}$ are the expansion coefficients of $p\cot\delta_{l}(p)$. For $l=0$, $a_{0}$ ($r_{0}$) is called the scattering length (effective range). At low energy, it is possible to neglect the expansion parameters except for the scattering length $a_{0}$, so the amplitude has the structure $f_{0}(p)\propto (-1/a_{0}-ip)^{-1}$. This shows that the pole at $p=0$ is simple, in accordance with the Jost function analysis. The eigenmomentum is found to be
\begin{align}
    p
    =&
    i/a_{0} 
    \label{eq:eigenmomentum} .
\end{align}
For positive (negative) $1/a_{0}$, the eigenmomentum is positive (negative) pure imaginary, which corresponds to the bound (virtual) state solution. To obtain the resonance solution above the threshold, the contribution from the negative effective range is needed~\cite{Hyodo:2013iga}. Even in this case, the low energy behavior $p\ll \sqrt{2/|a_{0}r_{0}|}$ is governed by Eq.~\eqref{eq:eigenmomentum}.\footnote{Here it is assumed that both $a_{0}$ and $r_{0}$ are finite. The case with infinitely large effective range will be discussed in Sec.~\ref{sec:decoupling}.} It is however worth noting that the valid region of Eq.~\eqref{eq:eigenmomentum} becomes small when $r_{0}$ is increased with a fixed $a_{0}$. In Sec.~\ref{sec:ereandtheorem}, it is shown that $r_{0}$ is large and negative when the bound state is dominated by the elementary component. This means that the relation $f_{0}(p)\propto (-1/a_{0}-ip)^{-1}$ breaks down at small $p$ for an elementary-dominated bound state. In this way, the size of the valid region of Eq.~\eqref{eq:eigenmomentum} reflects the structure of the bound state.

For $l\geq 1$ (in three dimensions), the effective range parameter cannot be neglected in the low energy because of the causality bound~\cite{Hammer:2010fw}. This is intuitively understood by the dominance of the $p^{2}$ term in comparison with the $ip^{2l+1}$ and other higher order terms. The low-energy amplitude behaves as $f_{l}(p)\propto (-1/a_{l}+r_{l}p^{2}/2)^{-1}$ so the pole at $p=0$ is double. This allows the direct transition from the bound state to the resonance for $l\neq 0$.

\section{Threshold behavior and field renormalization constant}\label{sec:Zrelation}

Here the threshold formula~\eqref{eq:bindingenergy} is derived from the nonlocal potential~\eqref{eq:effectiveint} by the expansion of Eq.~\eqref{eq:polecond}. Near the threshold, Eq.~\eqref{eq:polecond} is given by
\begin{align}
    E_{h}-\bar{M}_{0}-\delta M
    &= 
    \Sigma(E_{h})
    \label{eq:eigenvalue} .
\end{align}
It was found that $E_{h}$ is of the order of $\delta M$, so $E_{h}$ is regarded as a sufficiently small quantity. Expanding $\Sigma(E_{h})$ around $E_{h}=0$, a relation between $E_{h}$ and $\delta M$ is obtained as
\begin{align}
    E_{h}
    &= 
    \frac{1}{1-\Sigma^{\prime}(0)}\delta M ,
    \quad \Sigma^{\prime}(E)\equiv\frac{d\Sigma(E)}{dE} ,
    \label{eq:scaling}
\end{align}
in the leading order of $E_{h}$. The derivative of the self-energy is related to the field renormalization constant $Z$ which expresses the elementariness of the bound state~\cite{PR136.B816,Weinberg:1965zz,Hyodo:2011qc,Aceti:2012dd,Hyodo:2013nka}. The constant $Z$ is calculated from the relation of the channel coefficients,
\begin{align}
    \chi_{E_{h}}(\bm{q})
    \left(E_{h}-\frac{\bm{q}^{2}}{2\mu}\right)
    &= 
    c(E_{h})\bra{\psi_{0}}\hat{V}\ket{\bm{q}} ,
\end{align}
which is obtained from Eq.~\eqref{eq:Hamiltonian}. Because the wavefunction of the bound state is normalized, a relation holds for the summation of the wave functions,
\begin{align}
    |c(E_{h})|^{2}+
    \int |\chi_{E_{h}}(\bm{q})|^{2}d^{3}q
    &= 
    1 .
    \label{eq:normalization}
\end{align}
By using these relations, the field renormalization constant $Z(E_{h})$ is evaluated as the overlap of the bound state wave function with the purely bare state $\psi_{0}$ as
\begin{align}
    Z(E_{h})
    =
    \left|\bra{\Psi}
    \begin{pmatrix}
    \ket{\psi_{0}} \\
    0
    \end{pmatrix}\right|^{2}
    &= 
    |c(E_{h})|^{2}
    =
    \frac{1}{1-\Sigma^{\prime}(E_{h})} .
    \label{eq:Zdef}
\end{align}
It is shown that $Z$ takes the value $0\leq Z\leq 1$~\cite{Hyodo:2013nka}. Because of the normalization~\eqref{eq:normalization}, $1-Z=\int |\chi_{E_{h}}(\bm{q})|^{2}d^{3}q$ corresponds to the compositeness which expresses the probability of finding the scattering (two-body molecule) component in the bound state. Thus, in Eq.~\eqref{eq:scaling}, the leading contribution to $E_{h}$ from the shift of the bare mass $\delta M$ is given by the field renormalization constant at zero binding energy
\begin{align}
    E_{h}
    =Z(0)\delta M .
    \label{eq:Zrelation}
\end{align}
As will be shown in Sec.~\ref{sec:proof}, to have a pole at threshold for $l=0$, $Z(0)$ must vanish. Because this is a subtle problem, a detailed discussion for $Z(0)=0$ is presented in Sec.~\ref{sec:theorem}. In the present context, the vanishing of the field renormalization constant $Z(0)$ forbids the contribution proportional to $\delta M$. This ensures the $s$-wave scaling $E_{h}\propto \delta M^{2}$ in Eq.~\eqref{eq:bindingenergy}. 

For $l\neq 0$, $Z(0)$ expresses the elementariness of the zero energy bound state. When $Z(0)=1$, the bound state is regarded as a purely elementary state which is decoupled from the scattering channel. This is natural because the eigenvalue is given by $E_{h}=\delta M$ so that the scaling law of the bare mass is not modified by the threshold effect, as a consequence of the decoupling from the scattering channel. Comparison of Eq.~\eqref{eq:scaling} with the expansion of the Jost function leads to
\begin{align}
    \frac{\Sigma(0)}{1-\Sigma^{\prime}(0)}
    =-\frac{\alpha_{l}^{\prime}}{2\mu\beta_{l}} 
    \quad \text{for }l\neq 0  ,
\end{align}
which relates the self-energy and the expansion coefficients of the Jost function. 

It should be noted that the field renormalization constant is a model-dependent quantity. At first glance, however, one may think that $Z(0)$ for nonzero $l$ can be extracted from the hadron mass scaling near the threshold using Eq.~\eqref{eq:Zrelation}. This is unfortunately not the case, because the relation between the QCD parameter $x$ and the bare mass $\delta M$ inevitably specifies the basis to measure $Z(0)$. In other words, the definition of the bare \textit{hadron} mass $\delta M$ in QCD is model dependent.

\section{Compositeness theorem}\label{sec:theorem}

It is shown in Sec.~\ref{sec:Zrelation} that vanishing of the field renormalization constant is essential for the mass scaling in the $s$ wave. Here this ``compositeness theorem'' is proved. The statement is as follows.
\begin{quote}
\textit{If the $s$-wave scattering amplitude has a pole exactly at the threshold with a finite range interaction, then the field renormalization constant vanishes.}
\end{quote}
It is important to recall the different nature of the pole at the threshold for $l=0$ and for $l\neq 0$. The pole at the threshold is an ordinary bound state in the $l\neq 0$ case, while the $s$-wave pole represents the special state called zero energy resonance~\cite{Taylor}. It follows from the Schr\"odinger equation that the wave function at zero energy behaves as $1/r^{l}$ at large $r$. The wave function is therefore normalizable for $l\neq 0$, while with $l=0$ the wave function is not square integrable and does not represent a bound state. In this case, even with the finite range interaction, the wave function spreads to infinity. This is related with the divergence of the scattering length, which is essential for the low energy universality in few-body systems~\cite{Braaten:2004rn}. 

A naive interpretation of the theorem $Z(0)=0$ would be that the zero energy resonance is a purely composite state. However, a finite elementary component $|c(E_{h})|^{2}$ is not necessarily excluded from the wave function. In the $B\to 0$ limit, the wave function of the scattering state spreads to infinity. In this case, because of the normalization~\eqref{eq:normalization}, the \textit{fraction} of the finite elementary component is zero, in comparison with the infinitely large scattering component.\footnote{The usual normalization $\bra{\Psi}\kket{\Psi}=1$ is not applicable to the state vector with an infinite norm, such as the zero-energy resonance. The normalization of resonances is nevertheless ensured by the use of the Gamow vectors in the rigged Hilbert space~\cite{Gamow:1928zz,Bohm:1981pv}.} Thus, $Z(0)=0$ follows even with any finite admixture of the elementary component, because of the property of the scattering state. 

In the following, a proof of the theorem is first given for the nonlocal potential~\eqref{eq:effectiveint} in Sec.~\ref{sec:proof}. In Secs.~\ref{sec:ereandtheorem} and \ref{sec:pcandtheorem}, the theorem is shown to be valid for a general local potential, using the effective range expansion and the pole counting argument, respectively. 

It should be emphasized that the $B\to 0$ limit is qualitatively different from the finite $B$ case. For instance, $Z(B)=0$ with a finite $B$ implies the complete exclusion of the elementary component, because the scattering component is also finite. The structure of the bound state for finite $B$ is discussed in Sec.~\ref{sec:finite}. It is shown that for finite $B$, the value of $Z(B)$ is in principle arbitrary. The connection of the finite $B$ and $B\to 0$ limit becomes clear by considering the decoupling limit in Sec.~\ref{sec:decoupling}.

\subsection{Proof}\label{sec:proof}

Consider the field renormalization constant $Z$ of the bound state from the potential~\eqref{eq:effectiveint}. As shown in Eq.~\eqref{eq:Zdef}, $Z$ for the bound state with the binding energy $B=-E_{h}>0$ is related to the derivative of the self-energy as 
\begin{align}
    Z(B)
    =\frac{1}{1-\Sigma^{\prime}(-B)},
    \quad \Sigma^{\prime}(-B)=-\frac{d\Sigma(-B)}{dB}.
\end{align}
The $s$-wave self-energy~\eqref{eq:selfenergy} is given by
\begin{align}
    \Sigma(-B)
    =-4\pi\sqrt{2\mu^{3}}\int_{0}^{\infty}
    \frac{dE^{\prime}\sqrt{E^{\prime}}
    |F(E^{\prime})|^{2}}
    {E^{\prime}+B} ,
    \label{eq:selfenergyBfinite}
\end{align}
where the spherical $s$-wave form factor of the bare state is defined as $F(E^{\prime})=\bra{\psi_{0}}\hat{V}\ket{\bm{q}}$ with $E^{\prime}=|\bm{q}|^{2}/(2\mu)$. To reproduce the low energy limit of the scattering amplitude $f_{0}(p)\to (\text{const.})$ with Eq.~\eqref{eq:amplitude}, the factor $|F(E^{\prime})|^{2}$ should be an analytic function of the energy with a constant at small $E^{\prime}$.\footnote{The nonanalytic term $ip$ in the denominator of the amplitude comes from the imaginary part of the self-energy.} Thus, the factor is written as
\begin{align}
    |F(E^{\prime})|^{2}
    =g_{0}^{2}[1+\mathcal{O}(E^{\prime})] ,
\end{align}
where $g_{0}$ is the coupling constant of the bare state to the scattering state. First, examine the case where $g_{0}^{2}$ is nonzero in the limit $B\to 0$. The ultraviolet behavior of $|F(E^{\prime})|^{2}$ is also constrained to make the self-energy finite. Let $E_{\rm max}$ be the energy scale above which the integrand of Eq.~\eqref{eq:selfenergyBfinite} is sufficiently suppressed. With these conditions, the small $B$ behavior of the self-energy is extracted as 
\begin{align}
    \Sigma(-B)
    &\approx -4\pi\sqrt{2\mu^{3}}\int_{0}^{E_{\rm max}}
    \frac{dE^{\prime}\sqrt{E^{\prime}}
    g_{0}^{2}[1+\mathcal{O}(E^{\prime})]}
    {E^{\prime}+B} \nonumber \\
    &\propto 
    g_{0}^{2}
    \left[
    \sqrt{E_{\rm max}}
    -\sqrt{B}
    \arctan\left(\sqrt{\frac{E_{\rm max}}{B}}\right)
    +\dotsb \right]
    \nonumber \\
    &= 
    g_{0}^{2}
    \left[
    (\text{const.})
    +\mathcal{O}(B^{1/2})\right]
    \nonumber \\
    &\xrightarrow[B\to 0]{}
    (\text{finite}) .
    \label{eq:selfenergyB}
\end{align}
The derivative of the self-energy is calculated as
\begin{align}
    \Sigma^{\prime}(-B)
    &\propto 
    g_{0}^{2}
    \left[
    \frac{1}{\sqrt{B}}
    \arctan\left(\sqrt{\frac{E_{\rm max}}{B}}\right)
    +\dotsb 
    \right]\nonumber \\
    &= 
    g_{0}^{2}
    \left[
    \frac{\pi}{2\sqrt{B}}
    +\mathcal{O}(B^{0})\right]
    \nonumber \\
    &\xrightarrow[B\to 0]{}
    \infty .
    \label{eq:Sigmaderivative}
\end{align}
Thus, it is found that the field renormalization constant vanishes in the $B\to 0$ limit:
\begin{align}
    Z(B)
    =\frac{1}{1-\Sigma^{\prime}(-B)}
    &\xrightarrow[B\to 0]{}
    0 . \label{eq:result}
\end{align}
The divergence of the derivative of the self-energy at the threshold can also be shown by the spectral representation~\cite{Hyodo:2011js}. The essential point is that the term $\sqrt{B}\arctan(1/\sqrt{B})$ in Eq.~\eqref{eq:selfenergyB} below the threshold is a consequence of the analytic continuation of the imaginary part of the self-energy above the threshold. Because the imaginary part of the self-energy is constrained by the dispersion relation, Eq.~\eqref{eq:Sigmaderivative} always holds.

The only exception to the above argument is the case with $g_{0}^{2}\to 0$ in the $B\to 0$ limit where $\Sigma^{\prime}(0)$ and $Z(0)$ can be finite. However, the absence of the coupling to the scattering state also indicates that the bare state cannot affect the scattering amplitude. Namely, the scattering amplitude reduces to that for noninteracting particles which does not have a pole at the threshold. This contradicts the assumption of having a pole at $p=0$. 

Thus, $g_{0}^{2}$ must be nonzero and the theorem is proved by Eq.~\eqref{eq:result}. The $g_{0}\to 0$ limit will be further examined in Sec.~\ref{sec:decoupling} to discuss the structure of the bound state.

\subsection{Effective range expansion and composite theorem}\label{sec:ereandtheorem}

Next, consider the bound state from a general local potential, for which the result of Ref.~\cite{JMP1.319} is applicable. The scattering length and the effective range in the weak binding limit are expressed by the field renormalization constant and the binding energy as~\cite{Weinberg:1965zz}
\begin{align}
    a_{0}
    &=\frac{2(1-Z)}{2-Z}R ,\quad 
    r_{0}
    =\frac{-Z}{1-Z}R ,\quad 
    R
    =\frac{1}{\sqrt{2\mu B}} ,
    \label{eq:Weinberg}
\end{align}
where the correction terms of the order of the typical length scale of the interaction are neglected. As shown in Ref.~\cite{Hyodo:2013nka}, this formula provides the criteria to judge the structure of near-threshold bound state:
\begin{align}
    \begin{cases}
    a_{0}
    \ll -r_{0} & Z\sim 1, \text{(elementary dominance)}\; ,  \\
    a_{0}\sim R
    \gg r_{0} & Z\sim 0, \text{(composite dominance)}\; .
    \end{cases}\label{eq:criterion}
\end{align}

In the limit $B\to 0$, it follows that $R\to \infty$. If there is no constraint on the value of $Z(0)$, there are three possibilities:
\begin{align}
    \begin{cases}
    a_{0}=\infty,r_{0}=\text{(finite)} &: Z(0)=0 \\
    a_{0}=\infty,r_{0}=-\infty &: 0<Z(0)<1 \\
    a_{0}=\text{(finite)},r_{0}=-\infty  &: Z(0)=1
    \end{cases} .
\end{align}
For $0<Z(0)\leq 1$, the effective range should diverge. Intuitively, it is unlikely that the finite range interaction provides the infinitely large effective range. More rigorously speaking, $r_{0}=-\infty$ modifies the linear dependence of the eigenmomentum~\eqref{eq:eigenmomentum} into quadratic in $p$. This contradicts the fact that the pole at the $p=0$ is simple~\cite{JMP1.319}. Thus, only the case with $Z(0)=0$ can be realized. In this case, the composite dominance in Eq.~\eqref{eq:criterion} is always guaranteed by $a_{0}=\infty$ and finite $r_{0}$. It is emphasized again that this is only the dominance of the composite component, not the complete exclusion of the elementary component.

\subsection{Pole counting and composite theorem}
\label{sec:pcandtheorem}

The pole counting argument is also useful to understand the meaning of the theorem. Here the local potential is again considered. In Refs.~\cite{Morgan:1990ct,Morgan:1992ge}, the structure of the bound state is related to the pole positions in different Riemann sheets of the complex energy plane. For a given bound state pole, if there is a nearby pole in the different Riemann sheet (the shadow pole~\cite{Eden:1964zz}), then the bound state is dominated by the elementary component. This method is later related to the field renormalization constant~\cite{Baru:2003qq,Hyodo:2013iga}. The denominator of the effective range amplitude is a quadratic function of the eigenmomentum $p$. The pole positions can be analytically calculated as functions of $a_{0}$ and $r_{0}$. Using the relations~\eqref{eq:Weinberg}, they can be expressed by the binding energy and $Z$ as~\cite{Baru:2003qq}.
\begin{align}
    p_{1}=i\sqrt{2\mu B},
    \quad 
    p_{2}=
    -i\sqrt{2\mu B}
    \frac{2-Z}{Z} .
\end{align} 
The pole $p_{1}$ ($p_{2}$) is in the first (second) Riemann sheet in the energy plane and corresponds to the bound state (shadow) pole. For $Z\sim 1$ (elementary dominance), two poles have a similar energy $p_{1}^{2}/2\mu\sim p_{2}^{2}/2\mu$. For $Z\sim 0$ (composite dominance), the shadow pole $p_{2}$ goes away from $p_{1}$ and the bound state is essentially described by the pole $p_{1}$.

Now, consider the $B\to 0$ limit. If there is no constraint on the value of $Z(0)$, there are two possibilities:
\begin{align}
    \begin{cases}
     p_{1}=0, \quad p_{2}=-i(\text{finite}) &: Z(0)=0 \\
     p_{1}=p_{2}=0 &: 0<Z(0)\leq 1 
    \end{cases} .
\end{align}
In the $0<Z(0)\leq 1$ case, the pole at the threshold is double. This contradicts the simple pole at the $p=0$~\cite{JMP1.319}, and only the case with $Z(0)=0$ can be realized.

\subsection{Finite binding case}\label{sec:finite}

The above discussion is valid for the pole exactly at the threshold. This is an idealization of the physical hadronic states which have a finite binding energy $B\neq 0$. Here the bound state with a small but finite binding energy is examined.

For a given $B\neq 0$, it is always possible to tune the form factor $\bra{\psi_{0}}\hat{V}\ket{\bm{q}}$ and the bare mass $M_{0}$ such that the self-energy $\Sigma(-B)$ and its derivative $\Sigma^{\prime}(-B)$ take arbitrary values. In other words, the value of $Z(B)$ for $B\neq 0$ is in principle arbitrary. In the effective range expansion, for a finite scattering length, it is in principle possible to generate the effective range such that $a_{0}\ll -r_{0}$ which leads to the elementary dominance of the bound state.\footnote{After the submission of this paper, Ref.~\cite{Hanhart:2014ssa} appears on the web, which discusses the near-threshold scaling and its relation to the structure of the bound state. Reference~\cite{Hanhart:2014ssa} shows that the elementary dominance is realized by a ``significant fine tuning'', and it is natural to expect that the composite (molecular) state appears for small $B$.} It is only in the $B\to 0$ limit where the scattering length diverges and the nonzero $Z$ is forbidden.

It is instructive to compare the bound state case and resonance case. The arbitrariness of $Z$ for the bound state stems from the fact that the binding energy $B$ does not determine both $a_{0}$ and $r_{0}$. In contrast, because the pole position of a near-threshold resonance contains two independent quantities (real and imaginary parts), $a_{0}$ and $r_{0}$ are uniquely determined only by the pole position~\cite{Hyodo:2013iga}. What is missing in the bound state case is the position of the shadow pole in the second Riemann sheet. If the position of the shadow pole is given in addition to $B$, the field renormalization constant is uniquely determined for the bound state.

The weak binding formula~\eqref{eq:Weinberg} relates the field renormalization constant to the observables ($a_{0}$, $r_{0}$, and $B$). Because the observables do not depend on the specific model, it is sometimes mentioned that the structure of the weakly bound state is model-independently determined. Strictly speaking, to derive the weak binding formula~\eqref{eq:Weinberg} one implicitly assumes the absence of the singularity of the inverse amplitude [called the Castillejo-Dalitz-Dyson (CDD) pole~\cite{Castillejo:1956ed}] between the threshold and the bound state pole~\cite{Weinberg:1965zz}. Let $E=-C$ be the position of the closest CDD pole.\footnote{Thus $E=-B$ is the closest pole and $E=-C$ is the closest zero of the amplitude.} The effective range expansion breaks down at the singularity of the inverse amplitude closest to the threshold. Thus, if $-B<-C<0$, then the bound state pole locates outside of the valid region of the effective range expansion. In this case, the formula~\eqref{eq:Weinberg} is not applicable and the field renormalization constant cannot be related to the observables. On the other hand, when the effective range expansion is valid at the energy of the bound state pole ($-C<-B<0$), the field renormalization constant $Z$ can be related to the observables. Naively, having the CDD pole in the region $-B<E<0$ for a small $B$ requires a fine tuning, although there is no general principle to exclude this possibility.

\subsection{Decoupling limit}\label{sec:decoupling}

\begin{figure}[tbp]
    \centering
    \includegraphics[width=5cm,clip]{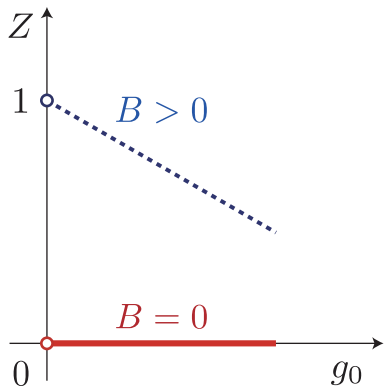}
    \caption{\label{fig:BZ}
    (Color online) Schematic illustration of the field renormalization constant $Z$ as a function of the coupling strength $g_{0}$ with a fixed binding energy $B$. Solid (dotted) line represents the $B=0$ ($B>0$) case. }
\end{figure}

The bound state pole disappears from the scattering amplitude in the $g_{0}\to 0$ limit, so this case is not relevant to the study of the mass scaling. Nevertheless, a detailed analysis of this decoupling limit provides an insight on the structure of the bound state. In Sec.~\ref{sec:proof}, the expression of $Z(B)$ for a small $B$ is found to be
\begin{align}
    Z(B)
    \approx\frac{1}{1-c\frac{g_{0}^{2}}{\sqrt{B}}} ,
\end{align}
where $c$ is a nonzero constant determined by kinematics. In the $g_{0}\to 0$ limit with a fixed $B>0$, the field renormalization constant $Z(B)$ behaves as
\begin{align}
    Z(B)
    \xrightarrow[g_{0}\to 0]{}
    1 \quad \text{for }B>0 .
    \label{eq:ZBg0}
\end{align}
This indicates that the bound state in this limit is a purely elementary state. Intuitively, the composite component disappears because of the absence of the coupling to the scattering state. If $g_{0}$ is decreased with a fixed $B>0$ with the potential~\eqref{eq:effectiveint}, the bare mass $M_{0}$ will approach the bound state pole position. In the $g_{0}\to 0$ limit, the bare pole locates exactly at $E=-B$, without the admixture of the scattering state. This is illustrated in Fig.~\ref{fig:BZ} (dotted line).

Although the scattering amplitude does not have the bound state pole, the bare state exists in the decoupled sector and is interpreted as an elementary particle. In other words, the purely elementary state with $Z=1$ cannot appear in the scattering amplitude by definition, because such a state does not have the scattering state component. Thus, the $Z=1$ state is realized only in the decoupled sector.

Next, consider the $g_{0}\to 0$ limit with $B=0$. As shown in Sec.~\ref{sec:proof}, $Z(0)$ is always zero for finite $g_{0}$. Thus, by taking the decoupling limit with keeping $B=0$, the field renormalization constant becomes
\begin{align}
    Z(0)
    \xrightarrow[g_{0}\to 0]{}
    0 \quad \text{for }B=0 .
    \label{eq:ZBB0}
\end{align}
This is also illustrated in Fig.~\ref{fig:BZ} (solid line). Through the comparison of this result with the $B\to 0$ limit of Eq.~\eqref{eq:ZBg0}, it is found that the two limits $B\to 0$ and $g_{0}\to 0$ do not commute with each other. Namely,
\begin{align}
    \lim_{B\to 0}\lim_{g_{0}\to 0}Z(B)
    =
    1 ,
\end{align}
while 
\begin{align}
    \lim_{g_{0}\to 0}\lim_{B\to 0}Z(B)
    =
    0 .
\end{align}
Thus, the value of $Z(B)$ at $B=g_{0}= 0$ is indefinite. In fact, in the simultaneous limit of $g_{0},B\to 0$, the value of $Z$ depends on how $g_{0}^{2}$ approaches zero:
\begin{align}
    \lim_{g_{0},B\to 0}Z(B)
    &=
    \begin{cases}
    0 & g_{0}^{2}\sim B^{1/2-\epsilon} \\
    \dfrac{1}{1-cD} & g_{0}^{2}\sim DB^{1/2} \\
    1 &  g_{0}^{2}\sim B^{1/2+\epsilon}
    \end{cases} 
\end{align}
with a positive $\epsilon$. 

The ambiguity of the limit value of $Z$ reflects the arbitrariness of $Z$ with finite $B$. As discussed in Sec.~\ref{sec:finite}, for $B>0$, the bound state with arbitrary $Z$ can be generated by tuning the model parameters such as $g_{0}$. During the $B\to 0$ process, the parameters can be continuously tuned such that the value of $Z$ remains the same. This eventually leads to $g_{0}\to 0$ in the $B\to 0$ limit, otherwise $Z=0$ should hold. Thus, to take the $B\to 0$ limit with keeping a finite $Z$, the bound state pole must disappear from the amplitude at the end. In this way, the state with a finite $Z$ can only be realized in the decoupled sector. To maintain the pole in the $B\to 0$ limit, $g_{0}$ must be kept finite and the field renormalization constant vanishes at the end.

\section{Model calculation}

It is illustrative to solve the eigenvalue equation by introducing a specific model for the interaction potential in the $l$th partial wave as
\begin{align}
    \bra{\bm{q}}\hat{V}\ket{\psi_{0}}
    &= 
    \bra{\psi_{0}}\hat{V}\ket{\bm{q}}
    = 
    g_{l}|\bm{q}|^{l}\Theta(\Lambda-|\bm{q}|)
    \label{eq:model} ,
\end{align}
with the real coupling constant $g_{l}$ and the cutoff parameter $\Lambda$. The $|\bm{q}|^{l}$ dependence is chosen to reproduce the low energy behavior of the amplitude $f_{l}(p)\sim p^{2l}$. The step function is introduced to tame the ultraviolet divergence. The self-energies for $l=0$ and $l=1$ channels are
\begin{align}
    \Sigma_{0}(E)
    &= -8\pi \mu g_{0}^{2} 
    \Biggl[\Lambda 
    -\sqrt{-2\mu E^{+}}
    \arctan\left(\frac{\Lambda}{\sqrt{-2\mu E^{+}}}\right)
    \Biggr] , \\
    \Sigma_{1}(E)
    &= 
    -8\pi \mu g_{1}^{2}
    \frac{\Lambda^{3}}{3}
    +2 \mu E \frac{g_{1}^{2}}{g_{0}^{2}}
    \Sigma_{0}(E) ,
\end{align}
where $E^{+}=E+i0^{+}$. The eigenvalue equation~\eqref{eq:eigenvalue} is numerically solved for these self-energies. For $\delta M<0$ ($\delta M>0$), the first (second) Riemann sheet of the complex energy plane is chosen to obtain the bound state (virtual and resonance state) solution. In this setup, the cutoff $\Lambda$ determines the scale of the system. The coupling constants are set to be $g_{0}^{2}=\Lambda/(100\mu^{2})$ and $g_{1}^{2}=1/(40\mu^{2}\Lambda)$. This leads to $\bar{M}_{0}\approx 0.25\Lambda^{2}/\mu$ for $l=0$ and $\bar{M}_{0}\approx 0.21 \Lambda^{2}/\mu$ for $l=1$.

\begin{figure*}[tbp]
    \centering
    \includegraphics[width=12cm]{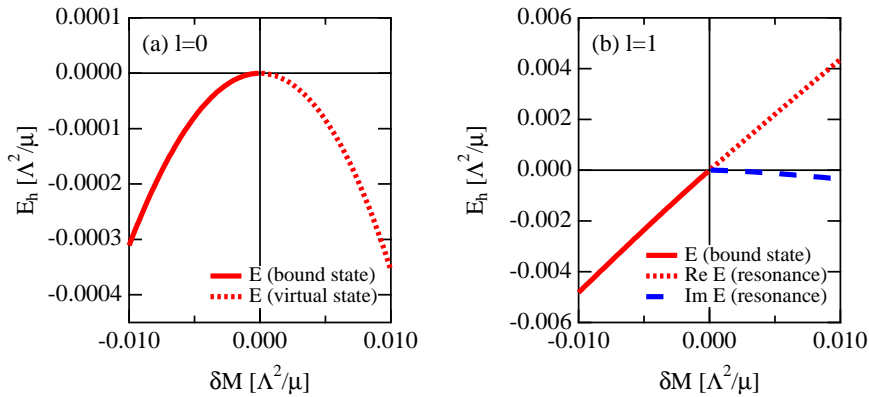}
    \caption{\label{fig:numerical1}
    (Color online) Near-threshold 
    eigenenergies as functions of $\delta M$ for $l=0$ (a) and for $l=1$ 
    (b). Solid, dotted, and dashed lines represent the energy in the 
    first Riemann sheet, the real part of the energy in the second 
    Riemann sheet, and the imaginary part of the energy in the second Riemann 
    sheet, respectively.}
\end{figure*}%

\begin{figure}[bp]
    \centering
    \includegraphics[width=6cm]{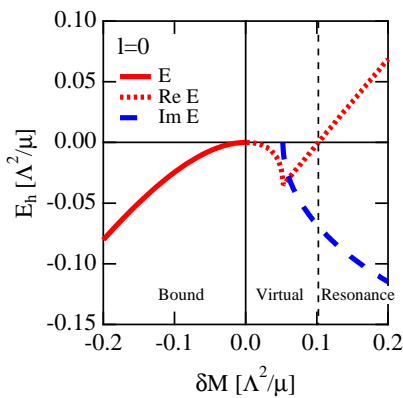}
    \caption{\label{fig:numerical2}
    (Color online) Eigenenergy for $l=0$ as a function of $\delta M$ in 
    the region $|\delta M |\leq 0.2\Lambda^{2}/\mu$. Solid, dotted, and 
    dashed lines represent the energy in the first Riemann sheet, the 
    real part of the energy in the second Riemann sheet, and the 
    imaginary part of the energy in the second Riemann sheet, 
    respectively.}
\end{figure}%

The near-threshold eigenenergies are shown in Figs.~\ref{fig:numerical1}(a) and \ref{fig:numerical1}(b). It is found that the near-threshold behavior follows the general scaling in Eqs.~\eqref{eq:bindingenergy} and \eqref{eq:eigenenergy}: quadratic dependence on $\delta M$ in the $s$ wave and linear dependence in the $p$ wave. As shown in Eq.~\eqref{eq:Zrelation}, the slope of the binding energy in the $p$-wave case is determined by the field renormalization constant at zero energy $Z(0)=[1-\Sigma_{1}^{\prime}(0)]^{-1}\approx 0.44$.

These behaviors are realized only near the threshold. If $\delta M$ is increased further, the virtual state in the $s$ wave acquires a finite width,\footnote{At the point where the imaginary part starts, the real part exhibits a cusp behavior. This nonanalytic cusp structure is essentially the same with what is discussed in Ref.~\cite{Guo:2012tg}.} and eventually goes above the threshold to become the resonance~\cite{Hyodo:2013iga}. This is demonstrated in Fig.~\ref{fig:numerical2}. 

It is clear from Fig.~\ref{fig:numerical2} that the scaling of the bound state energy is not continuously connected to the real part of the resonance energy near the $s$-wave threshold, because of the existence of the virtual state. This discontinuity is unavoidable, because it originates in the universal near-threshold scaling~\eqref{eq:bindingenergy} and \eqref{eq:eigenenergy}. The analysis with the effective range expansion shows that the energy region where the virtual state appears is determined essentially by the effective range parameter $r_{0}$. For instance, the deepest energy of the virtual state is $E_{h}=-1/(2\mu r_{0}^{2})$, and the width of the virtual state when it turns into the resonance is given by $\im E_{h}=-1/(\mu r_{0}^{2})$~\cite{Hyodo:2013iga}. This suggests that the size of the scaling violating region is determined by the inverse of the effective range parameter.

\section{Discussion}

\subsection{Chiral extrapolation}

The present result has an implication to the chiral extrapolation for the lattice QCD.\footnote{The present argument is based on the analyticity of the $S$ matrix which is not guaranteed in a finite volume where actual simulation is performed. The results in the infinite volume limit are considered.} In a naive application of chiral perturbation theory, the two-body loop effect is incorporated by perturbative calculations according to the systematic power counting. This corresponds to approximate Eq.~\eqref{eq:polecond} as 
\begin{align}
    E_{h}
    =&
    M_{0}+\Sigma(M_{0})+\dotsb .
    \label{eq:perturbative}
\end{align}
In this case, the scaling near the $s$-wave threshold becomes $E_{h}\propto \delta M$ and the universal result cannot be reproduced. It should be emphasized that the nonperturbative effect [self-consistent treatment in Eq.~\eqref{eq:polecond}] is essential for the universal behavior around the $s$-wave threshold. Indeed, inclusion of the nonperturbative dynamics through the dispersion relations~\cite{Hanhart:2008mx} shows the $m_{q}$ dependence consistent with the universal scaling. It is worth mentioning that the importance of the re-summation in chiral perturbation theory is known for the $NN$ scattering~\cite{Epelbaum:2008ga} and the $\bar{K}N$ scattering~\cite{Hyodo:2011ur}. A common feature for these sectors is the existence of the near-threshold $s$-wave (quasi) bound state, deuteron in the $NN$ scattering and $\Lambda(1405)$ in the $\bar{K}N$ scattering. One encounters the same situation during the mass scaling across the threshold, when the bound state pole approaches the $s$-wave threshold. Thus, the re-summation should be properly performed for the chiral extrapolation near an $s$-wave threshold.

In $p$ or higher partial waves, on the other hand, perturbative calculation~\eqref{eq:perturbative} provides an estimate of the field renormalization constant $Z(0)=[1-\Sigma^{\prime}(0)]^{-1}\approx 1+\Sigma^{\prime}(0)$, when the coupling of the bare state and the scattering state is small. The mass scaling for $l\neq 0$ can therefore be estimated by the usual perturbative calculation.

The present analysis shows that the mass of hadrons scales discontinuously near the $s$-wave threshold. This raises a caution on the use of the perturbative extrapolation formula when the physical state is expected to appear near the threshold. This problem may be avoided if one extrapolates the potential of the hadron-hadron interaction, which is continuous in $\delta M$, instead of the eigenenergy.

\subsection{Feshbach resonance of cold atoms}

The near-threshold behavior in the bound region is also studied for the Feshbach resonance in cold atom physics~\cite{Kohler:2006zz}. The energy of a shallow two-body bound state is proportional to the inverse scattering length squared $E_{2}\propto a_{0}^{-2}$, and the scattering length near a Feshbach resonance is given by $a_{0}(B^{\rm em})\propto [1-\Delta B^{\rm em}/(B^{\rm em}-B^{\rm em}_{0})]$ with the external magnetic field $B^{\rm em}$, its critical strength $B_{0}^{\rm em}$, and the width parameter $\Delta B^{\rm em}$~\cite{Kohler:2006zz}. The leading contribution to the binding energy is 
\begin{align}
    E_{2}
    &\propto 
    (B^{\rm em}-B_{0}^{\rm em})^{2}
    +\dotsb .
\end{align}
This shows the quadratic dependence of the binding energy on the strength of the magnetic field. Because the mass difference of the different spin states $\Delta M$ is proportional to $B^{\rm em}-B_{0}^{\rm em}$, the leading contribution to the binding energy is
\begin{align}
    E_{2}
    &\propto 
    (\Delta M)^{2}
    +\dotsb .
\end{align}
This is nothing but the scaling in Eq.~\eqref{eq:bindingenergy}. The field renormalization constant $Z$ at small binding energy is also calculated as~\cite{Braaten:2003sw,Duine:2003zz,Kohler:2006zz}
\begin{align}
    Z
    &\propto \frac{1}{a_{0}} 
    \propto \sqrt{|E_{2}|}
\end{align}
which is fully consistent with the compositeness theorem in Sec.~\ref{sec:theorem}. 

\subsection{Three-body bound state}

It is finally noted that the threshold scaling is universal for the \textit{two-body} bound state. It was found that the $s$-wave \textit{three-body} bound state directly turns into a resonance across the three-body breakup threshold when the Efimov effect occurs~\cite{Bringas:2004zz,Hyodo:2013zxa}. The three-body breakup process is beyond the applicability of the present framework. To analyze such behavior, it is needed to establish the low energy expansion of the three-body amplitude. The study of the scaling and compositeness of three-body bound states deserves an interesting future work.

\section{Summary}

The near-threshold behavior of the hadron mass scaling was is studied. By using the expansion of the Jost function, the general scaling law of the pole of the scattering amplitude is derived for a local potential. By utilizing the property of the field renormalization constant $Z$ in the zero binding limit, the same scaling is obtained for the nonlocal potential of Eq.~\eqref{eq:effectiveint}. It is shown for the $s$ wave that the scaling of the binding energy does not continuously connected to the real part of the resonance energy. 

A detailed discussion on the field renormalization constant of the zero energy resonance in the $s$ wave is presented. It is shown that, if there is a pole exactly at the threshold, the field renormalization constant should vanish. The vanishing of the field renormalization constant at zero energy guarantees the quadratic scaling of the binding energy in the $s$ wave. This result is interpreted as a consequence of the infinitely large two-body scattering component in the zero binding limit, which overwhelms any finite admixture of the elementary component. If one takes the zero binding limit with keeping finite $Z$, then the bound state pole decouples from the amplitude. 

The near-threshold scaling found here gives caution to the chiral extrapolation of the hadron mass across the $s$-wave threshold, because naive perturbative calculation does not reproduce the general scaling law. As in the case of the $NN$ and $\bar{K}N$ scattering in chiral perturbation theory, the nonperturbative re-summation is necessary to reproduce the correct threshold behavior.

\section*{Acknowledgments}

The author thanks Yusuke Nishida, Hideo Suganuma, and Sinya Aoki for fruitful discussions. This work is supported in part by JSPS KAKENHI Grants No. 24740152 and by the Yukawa International Program for Quark-Hadron 
Sciences (YIPQS).

\appendix
\section*{Appendix: Jost function}\label{sec:Jost}

Here the basic properties of the Jost function are summarized~\cite{Taylor}. The system to be considered in the following is the two-body elastic scattering by the spherical local potential $V(r)$ in the absence of the long-range force (such as the Coulomb interaction) so that the standard scattering theory can be formulated.

First, consider the regular solution of the Schr\"odinger equation $\phi_{l,p}(r)$ with the angular momentum $l$ and momentum $p$. This is the radial wave function normalized as $\phi_{l,p}(r)\to \hat{j}_{l}(pr)$ at $r\to 0$ with the Riccati-Bessel function $\hat{j}_{l}(pr)$. The regular solution follows the integral equation,
\begin{align}
    \phi_{l,p}(r)
    &= 
    \hat{j}(pr)+\int_{0}^{r}dr^{\prime}g_{l,p}(r,r^{\prime})U(r^{\prime})\phi_{l,p}(r^{\prime}) ,
    \tag{A1}
\end{align}
where $U(r)=2\mu V(r)$ and the free Green's function is given by $g_{l,p}(r,r^{\prime})=[\hat{j}_{l}(pr)\hat{n}_{l}(pr^{\prime})-\hat{n}_{l}(pr)\hat{j}_{l}(pr^{\prime})]/p$.

The Jost function $\Jost_{l}(p)$ is defined by the asymptotic behavior at $r\to \infty$ of the regular solution $\phi_{l,p}(r)$ as
\begin{align}
    \phi_{l,p}(r)
    &
    \xrightarrow[r\to \infty]{}
    \frac{i}{2}[\ \Jost_{l}(p)\hat{h}_{l}^{-}(pr)
    -\Jost_{l}(-p)\hat{h}_{l}^{+}(pr)] , 
    \tag{A2}
\end{align}
where $\hat{h}^{\pm}_{l}(z)=\hat{n}_{l}(z)\pm i\hat{j}_{l}(z)$ is the Riccati-Hankel function. The $s$ matrix $s_{l}(p)$ and the partial wave scattering amplitude $f_{l}(p)$ can be expressed by the Jost function as 
\begin{align}
    s_{l}(p)
    &= 
    \frac{\Jost_{l}(-p)}{\Jost_{l}(p)} ,\quad 
    f_{l}(p)
    = 
    \frac{\Jost_{l}(-p)-\Jost_{l}(p)}{2ip\Jost_{l}(p)} 
    \tag{A3}
\end{align}
Because the Jost function appears in the denominator, the zero of the Jost function is equivalent to the pole of the scattering amplitude.

From the comparison of the asymptotic form of the integral equation~(A1) with Eq.~(A2), the expression for the Jost function is obtained as
\begin{align}
    \Jost_{l}(p)
    &= 
    1+\frac{1}{p}
    \int_{0}^{\infty}dr \hat{h}_{l}^{+}(pr)
    U(r) \phi_{l,p}(r) .
    \tag{A4}
\end{align}
This is useful to expand the Jost function at small $p$. For $p\to 0$, the Riccati functions and the regular solution behave as
\begin{align}
    \hat{j}_{l}
    &\sim
    \phi_{l}\sim p^{l+1},\quad
    \hat{n}_{l}\sim p^{-l} .
    \tag{A5}
\end{align}
Thus, the expansion of the Jost function at small $p$ is given by
\begin{align}
    \Jost_{l}(p)
    &
    = 1+\alpha_{l}+\beta_{l}p^{2}+\mathcal{O}(p^{4})
    +i[\gamma_{l}p^{2l+1}+\mathcal{O}(p^{2l+3})] .
    \tag{A6}
\end{align}
The real expansion coefficients $\alpha_{l}, \beta_{l}, \gamma_{l},\dots $ depend on the potential $U$. 

Now, tune the potential $U$ such that the bound state appears exactly at the threshold. The condition to have a zero at $p=0$ is 
\begin{align}
    1+\alpha_{l}
    &
    = 0 .
    \tag{A7}
\end{align}
In this case, the expansion leads to 
\begin{align}
    \Jost_{l}(p)
    &
    = \beta_{l}p^{2}+\mathcal{O}(p^{4})
    +i[\gamma_{l}p^{2l+1}+\mathcal{O}(p^{2l+3})] ,
    \tag{A8}
\end{align}
which indicates Eq.~\eqref{eq:Jost}. In fact, the scaling~\eqref{eq:Jost} is shown on the general ground for a local potential~\cite{JMP1.319}, so that the leading coefficients $\gamma_{0}$ and $\beta_{l}$ ($l\neq 0$) cannot vanish. Next, introduce a small parameter $\delta \lambda$ to modify the potential as
\begin{align}
    U
    &
    \to (1+\delta\lambda)U .
    \tag{A9}
\end{align}
In this case, the expansion of the Jost function is given by
\begin{align}
    \Jost_{l}(p;\delta\lambda)
    &
    = 1+\alpha_{l}(\delta\lambda)+\beta_{l}(\delta\lambda)p^{2}
    +\mathcal{O}(p^{4}) \nonumber \\
    &\quad +i[\gamma_{l}(\delta\lambda)p^{2l+1}+\mathcal{O}(p^{2l+3})] ,
    \tag{A10}
\end{align}
with a condition $\alpha_{l}(0)=-1$. Expansion of the coefficients for small $\delta\lambda$ provides
\begin{align}
    \Jost_{l}(p;\delta\lambda)
    &= 
    \begin{cases}
    \alpha^{\prime}_{0}\delta\lambda 
    +i\gamma_{0}p + \mathcal{O}(p^{2},\delta\lambda p, \delta\lambda^{2}) & l=0 \\
    \alpha^{\prime}_{l}\delta\lambda 
    +\beta_{l}p^{2} + \mathcal{O}(p^{3},\delta\lambda p^{2},\delta\lambda^{2}) & l\neq 0 \\
    \end{cases} ,
    \tag{A11}
    \\
    \alpha^{\prime}_{l}
    &=\left.\frac{d\alpha_{l}}{d(\delta\lambda)}\right|_{\delta\lambda=0}
    ,\quad 
    \beta_{l}=\beta_{l}(0),\quad
    \gamma_{0}=\gamma_{0}(0),
    \tag{A12}
\end{align}
which leads to the eigenmomenta in Eqs.~\eqref{eq:momentum0} and \eqref{eq:momentuml}



\end{document}